\documentclass{iopconfser}

\usepackage{amsmath}
\usepackage{xcolor}
\usepackage{url}

\usepackage[sorting = none, backend = bibtex, style=numeric-comp]{biblatex}
\begin{document}

\title{Asymptotically safe quantum gravity: functional and lattice perspectives}

\author{Marc Schiffer$^{1}$}

\affil{$^1$ Institute for Mathematics, Astrophysics and Particle Physics, Radboud University \\ 
Heyendaalseweg 135, 6525 AJ Nijmegen, The Netherlands.}

\email{marc.schiffer@ru.nl}

\begin{abstract}
Asymptotically safe quantum gravity is a candidate theory to quantum gravity, which could unify the gravitational interaction with particle physics. It is characterized by quantum scale-symmetry at high energies. The constraining power of scale symmetry could be strong enough to even explain some parameters of the Standard Model of particle physics from first principles.
\end{abstract}

\section{Motivation: Quantum scale symmetry as fundamental symmetry of gravity and matter}
Formulating a unified description of gravity and matter, which is valid at any distance scale, is one of the biggest outstanding challenges in modern theoretical physics. 
On the one hand, the gravitational force in nature is accurately described by General Relativity (GR), which is tested through various distance scales, and in the strong-gravity regime. On the other hand, visible matter is accurately described in the Standard Model of particle physics (SM).
While both these ingredients describe our observations so far, they can both only be effective descriptions of nature: GR breaks down in the center of black holes, where curvature invariants are predicted to diverge. To resolve these divergences, quantum properties of spacetime itself, which are expected to become important below the Planck length, are necessary. The SM breaks down at tiny length scales, where some couplings are predicted to diverge in so-called Landau poles. The scale of these Landau poles is much below the Planck length, where quantum effects of spacetime become important, allowing for the appealing scenario where a unified theory of quantum gravity and matter solves both problems simultaneously, providing a fundamental description of nature.

One straightforward way to unify gravity and matter in one fundamental theory is to formulate both as a quantum field theory. However, quantizing GR perturbatively results in a loss of predictivity, since an infinite number of counter-terms needs to be added \cite{Goroff:1985th, vandeVen:1991gw}. Nevertheless, demanding that scale-symmetry is realized in the ultraviolet (UV), this problem can be resolved: in the same way that any symmetry constrains a theory, scale symmetry imposes infinitely many conditions on the couplings, and can therefore restore predictivity. This quantum realization of scale symmetry, which is called asymptotic safety, can be understood as a generalization of asymptotic freedom: in asymptotically free theories, all couplings are zero in the UV, hence restoring some form of classical scale invariance (e.g., in Yang-Mills theories). In asymptotically safe theories, the dimensionless versions of couplings are constant, but non-vanishing in the UV \cite{Weinberg:1980gg}. 
In the language of renormalization group flows, scale symmetry is realized at fixed points, where the coarse-graining-scale dependence of all couplings  vanishes. If this fixed point is realized in the UV, and at least one coupling is non-vanishing, it characterizes an asymptotically safe theory.

%
%
Here, we will summarize some key evidence for asymptotically safe quantum gravity (ASQG), which is most commonly explored with the Functional Renormalization Group (FRG), and Dynamical Triangulations (DT). We will also highlight mechanisms, how ASQG could cure Landau poles in the SM.
\section{Lattice and functional tools to explore asymptotically safe quantum gravity}
As an asymptotically safe theory is genuinely interacting in the UV, we need suitable tools to explore it. Early evidence for ASQG was obtained with perturbative tools in $d=2+\epsilon$ dimensions \cite{Gastmans:1977ad, Christensen:1978sc}, where the interacting fixed point arises from the free theory. Perturbative methods can also be employed in $d=4$ dimensions directly, where evidence for an interacting fixed point has been collected \cite{Niedermaier:2009zz, Niedermaier:2010zz, Falls:2024noj,  Kluth:2024lar}.
Besides these results, most evidence for this scenario comes from non-perturbative tools: continuum methods, in particular the FRG \cite{Percacci:2017fkn, Reuter:2019byg}, and lattice methods, in particular DT \cite{Loll:2019rdj}, provide independent evidence for the existence of a scale-invariant regime at high energies. Both methods require approximations and assumptions, resulting in systematic uncertainties of physical results. However, since both methods are genuinely different, their systematic uncertainties are uncorrelated. Hence, functional and lattice methods are \textit{complementary} tools to explore ASQG, in the same spirit as lattice and continuum methods complement each other, for example in studies of the strong interaction. Therefore, exploring properties of ASQG with both methods provides very robust evidence for those features.\\[-16pt]


\subsection{Evidence for asymptotic safety from functional methods}
\label{sec:FRG}
The key ingredient of the FRG is the scale-dependent effective average action $\Gamma_k$, where $k$ is the RG-scale. In comparison to the well-known $1$PI effective action $\Gamma$, $\Gamma_k$ only includes quantum fluctuations of modes with momenta $p^2>k^2$. Therefore, $\Gamma_k$ interpolates between an object akin to the classical action for $k\to\infty$ (no quantum fluctuations are integrated out), and the $1$PI effective action $\Gamma$ for $k\to0$ (all quantum fluctuations are integrated out). By lowering $k\to k-\delta k$, one can hence study the effect of modes with $p^2\approx k^2$ on the system. This change in the system is captured by a differential equation for $\Gamma_k$, the Wetterich equation \cite{Wetterich:1992yh, Morris:1993qb, Ellwanger:1993mw}, see \cite{Berges:2000ew, Pawlowski:2005xe, Gies:2006wv, Dupuis:2020fhh} for reviews.
The Wetterich equation requires Wick-rotation to Euclidean signature, as only there momentum scales can be used to clearly distinguish UV and IR modes. Recently, extensions to Lorentzian signature have been proposed \cite{Floerchinger:2011sc, Manrique:2011jc, Fehre:2021eob, DAngelo:2022vsh, DAngelo:2023wje, Saueressig:2025ypi}. These first studies suggest that Euclidean and Lorentzian results share important features on the qualitative level. 

In principle, $\Gamma_k$ contains all operators that are compatible with the symmetries of the theory. The flow equation can be used to extract the beta functions of couplings by projecting the flow on the desired operators. For gravity, there are infinitely many, diffeomorphism invariant operators, constructed from curvature scalars. Practical computations therefore  restrict $\Gamma_k$ to a subset of operators, which constitutes an approximation of the dynamics of the system. This restriction introduces systematic uncertainties in the results. These can be reduced by increasing the set of included operators, and assessed, for example, by studying residual gauge and regulator dependencies, see, e.g., \cite{Falls:2013bv, deBrito:2023myf, Baldazzi:2023pep, Riabokon:2025ozw}.
For gravity, the simplest approximation involves the Einstein-Hilbert action only,
parameterized by the couplings $G_k$ and $\lambda_k$, corresponding to the dimensionless versions of the Newton coupling and the cosmological constant, respectively.
Their beta-functions can be extracted from the Wetterich equation, which for $G_k$ schematically reads:
\begin{equation}
\beta_{G_k}=2\,G_k + \#\,G_k^2 +\mathcal{O}(G_k^3)\,,
\label{eq:betagschem}
\end{equation}
where the symbol $\#$ captures the effect from quantum fluctuations of spacetime. The first term in \eqref{eq:betagschem} arises from the canonical scaling of the Newton coupling. Therefore, asymptotic safety, which is characterized by $G_*>0$ for which $\beta_{G_k}|_{G_*}=0$ holds, arises from a balancing of canonical and quantum scaling, if $\#>0$.
Starting with the pioneering work \cite{Reuter:1996cp}, compelling evidence for the existence of the asymptotically safe, so-called Reuter fixed point, has been collected in a large range of approximations and setups, see \cite{Percacci:2017fkn, Reuter:2019byg, Bonanno:2020bil, Morris:2022btf, Martini:2022sll, Knorr:2022dsx, Eichhorn:2022bgu, Platania:2023srt, Eichhorn:2022gku, Pawlowski:2023gym} for introductions and overviews. There is also evidence that the fixed point is near perturbative \cite{Falls:2013bv, Falls:2014tra, Falls:2017lst, Falls:2018ylp, Eichhorn:2018akn, Eichhorn:2018ydy, Eichhorn:2018nda, Kluth:2020bdv} characterized by three free parameters only, highlighting the predictive power of scale symmetry \cite{Falls:2014tra, Denz:2016qks, Falls:2017lst, Falls:2020qhj}. This implies that approximations of $\Gamma_k$ based on canonical power-counting, can produce qualitatively reliable results. Long-standing questions regarding unitarity, the absence of unphysical modes, and Lorentzian signature have been addressed recently \cite{Knorr:2019atm, Platania:2020knd, Fehre:2021eob, Knorr:2021niv, DAngelo:2022vsh, Platania:2022gtt, DAngelo:2022vsh, DAngelo:2023wje, Pastor-Gutierrez:2024sbt,   Saueressig:2025ypi}.

Intriguingly, there is also strong evidence that the Reuter fixed point exists in the presence of SM matter \cite{Dona:2013qba, Meibohm:2015twa, Biemans:2017zca, Alkofer:2018fxj, Wetterich:2019zdo, Korver:2024sam}. Conversely, there is evidence that this fixed point induces a UV-completion in the matter sector, and might even predict some SM-parameters, see \cite{Shaposhnikov:2009pv, Harst:2011zx, Eichhorn:2017ylw, Eichhorn:2017lry, Eichhorn:2018whv, Alkofer:2020vtb, Kowalska:2022ypk, Pastor-Gutierrez:2022nki, Eichhorn:2025sux}: focusing on the Abelian gauge sector, the scale dependence of the Abelian hypercharge $g_{y}$, under the impact of ASQG schematically reads
\begin{equation}
    \beta_{g_{y}}=-f_{g}\,g_{y}+ \#_{\mathrm{matter}}\,g_y^{3}+\mathcal{O}(g_{y}^{4})\,
\end{equation}
where $\#_{\mathrm{matter}}>0$ is the pure-matter contribution, and where $f_g$ is the gravitational contribution. As $f_g$ is parameterized by the gravitational couplings, it is constant above the Planck scale, and quickly drops to zero below. In the SM, the screening property of charged matter (encoded in $\#_{\mathrm{matter}}>0$) implies an IR-attractive fixed point $g_{y,\,*}\!=0$, which results in the triviality problem. There is strong evidence that ASQG cures this problem with an anti-screening gravitational contribution $f_g>0$ \cite{Daum:2009dn, Harst:2011zx, Folkerts:2011jz, Christiansen:2017gtg, Eichhorn:2017lry, Christiansen:2017cxa, Eichhorn:2019yzm}.\footnote{Note that $f_g=0$ is also possible for specific regulators \cite{Folkerts:2011jz}. Then ASQG would not cure the Landau pole problem.} This contribution turns the free fixed point into a UV-relevant direction, inducing asymptotic freedom. $f_g>0$ gives also rise to an additional fixed point $g_{y,\,*}>0$, from which the low-energy value of the Abelian hypercharge can be predicted, see \cite{Harst:2011zx, Eichhorn:2017lry}. This mechanism highlights how ASQG could UV complete the SM, and even predict some SM couplings. Similar mechanisms are present in other sectors of the SM, indicating that ASQG provides a UV-completion for the SM \cite{Alkofer:2020vtb, Kowalska:2022ypk, Pastor-Gutierrez:2022nki}, with enhanced predictive power \cite{Shaposhnikov:2009pv, Eichhorn:2018whv, Eichhorn:2025sux}.\\[-16pt]

\subsection{Evidence for asymptotic safety from lattice methods}
Lattice methods are well-established tools to explore (quantum-) field theories in non-perturbative regimes. When applying these to gravity, the lattice itself must become dynamical, as \textit{any} regular, fixed,  lattice would constitute a preferred background, and hence break coordinate invariance. One possibility to formulate GR without coordinates is to discretize spacetime into triangles and the resulting higher-dimensional simplices \cite{Regge:1961px}. In terms of such a triangulation, the Einstein-Hilbert action can be translated into a discretized version, the so-called Einstein-Regge action $S_{\mathrm{ER}}$. In the simplest form in $d=4$ dimensions, it depends on the total number of triangles and hyper-tetrahedra, and is parameterized by two lattice couplings, related to the Newton coupling and the cosmological constant.
With this discretization, the path integral over the metric is replaced by a sum over all possible triangulations $T$.

In practice, the sum over triangulations is performed via Monte-Carlo sampling, where different triangulations are connected by Pachner-moves \cite{PACHNER1991129}, which ensure ergodicity. ASQG then corresponds to a continuous phase transition in the lattice-parameter space: there, the correlation length diverges, such that the lattice loses any scale, and hence becomes scale invariant. Since the loss of scales also means that the lattice spacing can be safely removed, a continuous phase transition is a necessary requirement to recover a continuum limit, see \cite{Loll:2019rdj, Ambjorn:2022naa, Budd:2022zry, Durhuus:2022rcb,Watabiki:2022aaj, Sato:2022ory, Clemente:2023sft, Gorlich:2023wtg,Loll:2023hen, Benedetti:2022ots, Gizbert-Studnicki:2023usv,} for recent reviews.\footnote{\cite{Ambjorn:2024qoe} argues that for gravity, the infinite volume limit automatically implies a diverging correlation length. Following this argument, a continuum limit can be taken anywhere in the parameter space.}
In $d=2$ (Euclidean) dimension, the described procedure does indeed yield a continuous phase transition, reproducing the critical exponents of two-dimensional bosonic strings, see, e.g., \cite{Jurkiewicz:1985sz, Ambjorn:1985dn, Polyakov:1981rd,}.
However, in $d=4$ dimensions, this formulation of DT does not feature a suitable phase transition, nor any phase that resembles smooth geometries \cite{Ambjorn:1993sy, Catterall:1995aj, deBakker:1994zf, Ambjorn:1995dj}.\\[-6pt]

This problem can be resolved by starting from the Lorentzian path integral and performing a proper analytical continuation, which is called Causal Dynamical Triangulations (CDT) \cite{Ambjorn:1998xu, Ambjorn:1999nc, Ambjorn:2001cv}. This results in a sum over only those Euclidean geometries, which have a Lorentzian continuation. In practice, this is implemented by distinguishing spacelike and timelike edges of a triangle and by restricting the way in which theey can be glued together. The Regge action is then slightly modified and includes an additional parameter controlling the difference between spacelike and timelike edge-lengths. This implementation of causal structure results in a rich phase-diagram with several first- and second-order phase transitions \cite{Ambjorn:2004qm, Ambjorn:2005qt, Ambjorn:2010hu, Ambjorn:2011cg,}. Importantly, the theory also features a phase where geometries feature a fractal, so-called Hausdorff dimension of $d_{H}=4$ \cite{Ambjorn:2012jv}, and are de Sitter like in the sense that the spatial volume-profiles, when averaged over many configuration, resemble that of de Sitter space \cite{Ambjorn:2007jv, Ambjorn:2008wc}.

While in gravity observables cannot be local due to diffeomorphism invariance, observables that are expected to capture some local features have been explored, which include the \textit{quantum Ricci curvature} \cite{Klitgaard:2017ebu, Klitgaard:2018snm, Klitgaard:2020rtv, Brunekreef:2020bwj,} and curvature-curvature correlators \cite{vanderDuin:2024pxb, Maas:2025rug}. These objects are fully coordinate-independent observables that capture non-perturbative aspects of quantum scale symmetry. They are therefore suitable quantities to compare with functional methods, but also to compare physical features of ASQG with other approaches to quantum gravity. Specifically, similar observables have been used to apply a RG procedure on the lattice, finding indications for an interacting RG fixed point \cite{Ambjorn:2024qoe, Ambjorn:2024bud}. Hence, lattice simulations provide some evidence for the Reuter fixed point, albeit not fully conclusive to date.\\[-6pt]

Besides the generalization of DT to include causal structure, a different modification has been considered in \cite{Bruegmann:1992jk, Laiho:2011ya, Ambjorn:2013eha, Laiho:2016nlp}, where an additional term is included in the partition function. This term, when considered as a part of the classical action, can be thought of as including some higher-order curvature invariants. If such an operator becomes UV-relevant, as indicated by FRG studies, it has to be tuned on the lattice and hence included in the classical action. Extending the parameter space in this way still results in only two phases separated by a first-order transition, which is not suitable for taking a continuum limit \cite{Ambjorn:2013eha, Laiho:2016nlp}. However, following along this phase transition, the geometries do become more de Sitter like \cite{Laiho:2016nlp, Dai:2024vjc}, and might feature well-behaved semi-classical and non-relativistic limits \cite{Dai:2021fqb, Bassler:2021pzt}. Furthermore, the lattice spacing decreases along the first-order line, giving hope for a critical endpoint \cite{Dai:2024vjc}. Therefore, also the inclusion of an additional relevant direction into the partition function might give rise to a well-behaved continuum limit, characterized by quantum scale symmetry.

\section{Summary and Outlook}

Asqymptotically safe quantum gravity is a candidate theory that unifies gravity and particle physics into one quantum field theory governed by scale symmetry. Such a scale-invariant regime, realized at an interacting RG fixed point in the UV, can be explored with different methods. Here, we focussed on functional and lattice methods, which are complementary to each other. Both methods involve approximations and assumptions, which lead to systematic uncertainties. On the functional side, these approximations are related to approximations on the dynamics of the system, encoded in the ansatz for $\Gamma_k$, which result in systematic uncertainties of scaling exponents or scattering amplitudes. On the lattice side, the approximations are related to finite lattice spacings and finite volumes, and the resulting extrapolations. Hence, the systematic uncertainties of both methods are expected to be independent, such that both methods are complementary tools to explore a scale-symmetric regime of quantum gravity and matter. The fact that both methods individually find evidence for an interacting RG fixed point can be regarded as strong evidence that this fixed point indeed exists, and is not an artifact of approximations.

In the same spirit, investigating gravity-matter systems on the lattice could confirm, if the qualitative mechanisms to UV-complete SM sectors, as discovered with FRG-techniques, see Section~\ref{sec:FRG}, indeed are robust physical features. Such a confirmation would significantly increase the confidence in the robustness of those mechanisms. First insights into this question can already be obtained in a quenched approximation, where the back-reaction of matter on the geometry is neglected. Furthermore, employing both methods in a concerted fashion, for example to extract scaling exponents of matter couplings, would allow to elevate the qualitative mechanisms explored so far to robust, quantitatively reliable features.

\section*{Acknowledgements}
MS thanks Renate Loll and Frank Saueressig for comments on the manuscript. This research was in parts supported by a Radboud Excellence fellowship from Radboud University in Nijmegen, Netherlands, and by an NWO Veni grant under grant ID  [\url{https://doi.org/10.61686/SUPEH07195}]. MS also acknowledges financial support by the COST Action CA23130 (“Bridging high and low energies in search of quantum gravity (BridgeQG)”) to attend the GR24 conference.


\printbibliography
\end{document}